\documentclass[largeformat]{interact}

\usepackage{epstopdf}
\usepackage[caption=false]{subfig}

\usepackage[numbers,sort&compress,merge]{natbib}
\bibpunct[, ]{[}{]}{,}{n}{,}{,}
\renewcommand\bibfont{\fontsize{10}{12}\selectfont}

\theoremstyle{plain}

\theoremstyle{definition}

\theoremstyle{remark}

\usepackage{siunitx}
\begin{document}

\articletype{ARTICLE SPECIAL ISSUE}

\title{Spin-orbit-dependent lifetimes of long-range Rydberg molecules}
\author{
\name{
Michael Peper\textsuperscript{$^\ast$}\thanks{$^\ast$) Present address: Department of Electrical and Computer Engineering, Princeton University, Princeton, New Jersey 08544, USA},
Jakob Skrotzki,
Martin Trautmann\textsuperscript{$\dagger$}\thanks{$\dagger$) Present address: Institute of Physics, Martin-Luther-University Halle-Wittenberg, Von-Danckelmann-Platz 3, 06120 Halle (Saale)},
Ajith Sanjai C. V. Sivakumar, 
and Johannes Deiglmayr\textsuperscript{$\ddagger$}\thanks{$\ddagger$) CONTACT J. Deiglmayr, Email: johannes.deiglmayr@physik.uni-leipzig.de}}
\affil{Felix-Bloch Institute, Leipzig University, Linn{\'e}stra{\ss}e 5, 04103 Leipzig, Germany }
}

\maketitle

\begin{abstract}
Long-range Rydberg molecules (LRMs) form when a highly excited Rydberg electron scatters from ground-state atoms inside its orbit, creating oscillatory, long-range potentials. We present a combined theoretical and experimental study of caesium dimers correlated to $40^2P_{3/2}$ Rydberg states, with an emphasis on decay via autoionisation (associative ionisation). Our model includes a relativistic treatment of electron–atom scattering with spin–orbit coupling, the perturber’s hyperfine structure, and coupling of vibrational levels to a continuum of short-range decay channels. Calculated potential-energy curves predict two families of wells: outer wells near the classical outer turning point supporting long-lived states, and inner wells at shorter range whose lifetimes are limited by tunneling and subsequent vibronic decay. Using photoassociation in an ultracold Cs gas and an analysis of pulsed-field-ionisation signals which are highly selective for the detection of molecules, we assign resonances by binding energy and measure lifetimes. The measured lifetimes of inner-well states increase systematically with increasing detuning and agree with calculated lifetimes; detection of Cs$_2^+$ product ions supports autoionisation as a dominant channel. We show that the lifetimes are strongly reduced by spin–orbit interactions in the transient Cs$^-$ collision complex, which lift the near-degeneracy in $\Omega$ observed for states in the outer well and control the inner-well binding. The identified states also provide promising pathways to create ultracold molecules in ion-pair states.
\end{abstract}


\section{Introduction}\label{sec:introduction}

Long-range Rydberg molecules (LRMs) are molecular bound states in which one constituent atom is in a highly excited Rydberg state and the remaining constituent atoms are in the ground state~\cite{eilesTrilobitesButterfliesOther2019,feyUltralongrangeRydbergMolecules2020,shafferUltracoldRydbergMolecules2018}. Binding arises from an effective Fermi-contact interaction between the almost-free Rydberg electron and ground-state perturbers within its orbit, characterised to leading order by the elastic $s$-wave scattering length $a$~\cite{fermiSopraSpostamentoPressione1934,masnou-seeuwsModelPotentialCalculations1982,kleimenovSpectroscopicCharacterizationPotential2008,hoganNewPerspectiveBinding2009a}. The oscillatory spatial probability density distribution of the electron in a Rydberg state of high principal quantum number $n$ leads to oscillatory potential energy curves that can support vibrational bound molecules with bond lengths $R$ corresponding to the classical outer turning point of the Rydberg electron orbit at $R=2 a_0 n^2$~\cite{greeneCreationPolarNonpolar2000}. Experimentally, LRMs were first observed by photoassociation spectroscopy in an ultracold, dense cloud of rubidium atoms~\cite{bendkowskyObservationUltralongrangeRydberg2009}. Since then, LRMs of homonuclear dimers, trimers and other small oligomers of rubidium~\cite{bendkowskyRydbergTrimersExcited2010}, 
caesium~\cite{tallantObservationBlueshiftedUltralongRange2012}, strontium~\cite{desalvoUltralongrangeRydbergMolecules2015}, and potassium~\cite{peperPhotodissociationLongrangeRydberg2020}, as well as heteronuclear dimers~\cite{whalenHeteronuclearRydbergMolecules2020,peperHeteronuclearLongRangeRydberg2021,guttridgeIndividualAssemblyTwoSpecies2025} have been experimentally studied. Perturbers with a more complex, multi-channel structure have been investigated theoretically~\cite{wojciechowskaUltralongrangeRydbergMolecules2025,eilesUltracoldLongRangeRydberg2015}. 

LRMs are a very unusual molecular system where the typical order of energy scales, from the largest electronic contribution down to the hyperfine structure, is reversed~\cite{andersonPhotoassociationLongRangeND2014,sassmannshausenExperimentalCharacterizationSinglet2015}, vibronic states of homonuclear dimers may exhibit very large dipole moments~\cite{boothProductionTrilobiteRydberg2015, niederprumObservationPendularButterfly2016}, and non-adiabatic effects are expected to play a large role~\cite{hummelSyntheticDimensionInducedConical2021,durstNonadiabaticCouplingsStabilization2025}. They are formed in electronically highly excited states and thus are subject to a number of decay processes (exemplified in the following for a dimer $X_2$): \textit{i)} radiative decay to a pair of unbound ground-state atoms $X+X$~\cite{bendkowskyObservationUltralongrangeRydberg2009}; \textit{ii)} black-body induced transitions to a Rydberg atom in a different state and an unbound ground-state atom $X^*+X$ ~\cite{butscherLifetimesUltralongrangeRydberg2011}; \textit{iii)} black-body induced ionisation, creating an atomic ion and an unbound ground-state atom $X^+ + X + e^-$; \textit{iv)} self-dissociation of the molecule through vibronic couplings into a Rydberg atom in a lower electronic state and an unbound ground-state atom $X^{*'}+X$, also called a ``state-changing collision''~\cite{schlagmullerUltracoldChemicalReactions2016,geppertDiffusivelikeRedistributionStatechanging2021}; \textit{v)} autoionisation through vibronic couplings into a stable bound molecular ion $X_2^+ + e^-$~\cite{schlagmullerUltracoldChemicalReactions2016}. Depending on the vibronic state of the LRM, all of these processes can play a role. 

In this article, we specifically investigate decay via the last mechanism above, which has also been referenced as associative ionisation~\cite{schlagmullerUltracoldChemicalReactions2016,niederprumGiantCrossSection2015}, or Hornbeck-Molnar ionisation~\cite{bendkowskyRydbergTrimersExcited2010}. We present the results of a theoretical model for Cs$_2$ LRMs correlated to $n^2P_{3/2}$ Rydberg states which includes the relativistic treatment of the elastic electron-perturber interaction, the hyperfine structure of the perturber, and the coupling of vibrational levels to a continuum of decay channels. Photoassociation resonances in an ultracold gas of caesium atoms are assigned to the calculated states based on binding energies. The assignment is supported by measurements of the lifetimes of the formed LRMs, which are in good agreement with calculated values. We demonstrate that the lifetimes of the studied LRMs are strongly influenced by the spin-orbit interaction in the Cs$^-$ collision complex, connecting to previous work on autoionisation rates of Rydberg molecules with channel interactions mediated by the ionic core~\cite{lefebvre-brionTheoreticalStudySpinorbit1985,wornerRoleSpinsMolecular2007}

\section{Vibronic structure of the long-range Rydberg states}\label{sec:theory}

\paragraph*{Electronic structure}

The Fermi-contact interaction~\cite{fermiSopraSpostamentoPressione1934}, extended to non-vanishing collision energies by Omont~\cite{omontTheoryCollisionsAtoms1977}, of a Rydberg atom in a state with quantum numbers $n,l,j,m_j$ and a spatial wavefunction $\Psi_{nljm_j}$, separated by the distance $\vec{R}$ from a ground-state atom, takes the following form~\cite{khuskivadzeAdiabaticEnergyLevels2002}:
\begin{equation}\label{eq:fermicontact}
    V_{\mathrm{FC}}(R) = 2 \pi a_S\left(k(R)\right) \left|\Psi_{nljm_j}(\vec{R})\right|^2 + 6 \pi a_P\left(k(R)\right) \left|\nabla\Psi_{nljm_j}(\vec{R})\right|^2 + \dots ,
\end{equation}
where $a_S(k)$ and $a_P(k)$ are the energy-dependent elastic electron-atom $s$-wave scattering length and $p$-wave scattering volume, respectively. The relevant collision energy is the semi-classical kinetic energy of the Rydberg-electron at distance $R$ from the ionic Rydberg core~\cite{greeneCreationPolarNonpolar2000}, 
\begin{equation}\label{eq:semiclassk}
    \frac{k^2}{2} = -\frac{1}{2{n^*}^2} + 1/R,
\end{equation}
where $n^*$ is the effective quantum number $n^*=n-\delta_{nlj}$ of the Rydberg state with quantum defect $\delta_{nlj}$. We constrain $k$ to a minimum value of \num{1e-4} to avoid the divergence of the $p$-wave contributions as $k\rightarrow0$. Note that atomic units are used everywhere, except if noted otherwise. 

To calculate the potential-energy curves resulting from Eq.~\eqref{eq:fermicontact}, we follow Ref.~\cite{eilesHamiltonianInclusionSpin2017}. The total Hamiltonian
\begin{equation}\label{eq:isoelectronic}
   H=H_{0}+H_\mathrm{HF} + V_\mathrm{FC},
\end{equation}
including the energies of the isolated Rydberg atom with its fine structure, $H_{0}$, the hyperfine interaction in the ground-state atom, $H_\mathrm{HF}$, and the Fermi-contact interaction $V_\mathrm{FC}$, is expressed through its matrix elements $\bar{H}$ in a basis composed of atomic Rydberg states and the electron- and nuclear-spin states of the ground-state atom, $\left\{nljm_j,m_sm_i\right\}$. The coupling of angular momenta is detailed in Appendix~\ref{sec:couplingScheme}. The Hamiltonian~\eqref{eq:isoelectronic} conserves the projection of the total electronic angular momentum on the internuclear axis, $\Omega=m_j+m_i+m_s$. We therefore compute and diagonalise $\bar{H}$  independently for each value of $\Omega$. The dependence of the electron-neutral scattering phase shifts on the quantum numbers of the electron-neutral collision complex $L, S, J, M_J$ is included through a frame transformation. $R$-dependent eigenvalues are then obtained by diagonalisation of the matrix $\bar{H}$ for different $R$. 

Due to the delta-function character of the Fermi-contact interaction, the eigenvalues of the matrix $\bar{H}$ do not converge when increasing the basis size~\cite{feyComparativeAnalysisBinding2015}. Nevertheless, comparison with more accurate Green's function calculations~\cite{greeneGreensfunctionTreatmentRydberg2023} has shown that the diagonalisation approach employed here can reach spectroscopic accuracy for LRMs correlated to a Rydberg state with angular momentum $l\leq2$ if the basis $\left\{nljm_j\right\}$ is chosen to include one Rydberg manifold (\textit{i.e.}, all Rydberg states for a given $n$ with $l>4$) below and one above the low-$l$ state of interest~\cite{peperHeteronuclearLongRangeRydberg2021,guttridgeIndividualAssemblyTwoSpecies2025}. The scattering parameters $a_{L,S,J}$ entering in the Hamiltonian~\eqref{eq:isoelectronic} must therefore be considered as effective parameters modeling the bound states of LRMs and are not only determined by the free-electron--atom scattering phase shifts. However, previous experimental studies demonstrated that an optimised set of scattering parameters describes the binding energies of low-$l$ LRMs over a large range of $n$-values if the closest-lying atomic Rydberg state is chosen as reference state $n^*$ in Eq.~\eqref{eq:semiclassk}~\cite{peperHeteronuclearLongRangeRydberg2021}.

We obtain the energy-dependent electron-caesium scattering parameters $a_{L,S,J}$ by following the relativistic model-potential method of Khuskivadze \textit{et al.}~\cite{khuskivadzeAdiabaticEnergyLevels2002}. We numerically solve the coupled differential equations for a given collision energy using the renormalised Numerov method~\cite{johnsonRenormalisedNumerovMethod1978} on a uniform grid of distances, starting from $r=0.014\,a_0$, with the initial values derived in Bahrim \textit{et al.} \cite{bahrimBoundaryConditionsPauli2001} for $^3P_J$, and Khuskivadze \textit{et al.} \cite{khuskivadzeAdiabaticEnergyLevels2002} for $S$- and $^1P_1$-wave channels. Phase shifts are then extracted by matching the solutions at $r_\mathrm{end}=500\,a_0$ to spherical Bessel functions. To improve the agreement with experimental binding energies, we make slight adjustments to the parameters of the model potentials from the values given by Khuskivadze \textit{et al.} \cite{khuskivadzeAdiabaticEnergyLevels2002}. Details on the computational implementation have been given in previous work \cite{peperPrecisionSpectroscopyUltracold2020,peperHeteronuclearLongRangeRydberg2021}. 

\begin{figure}
\centering
\resizebox*{7cm}{!}{\includegraphics{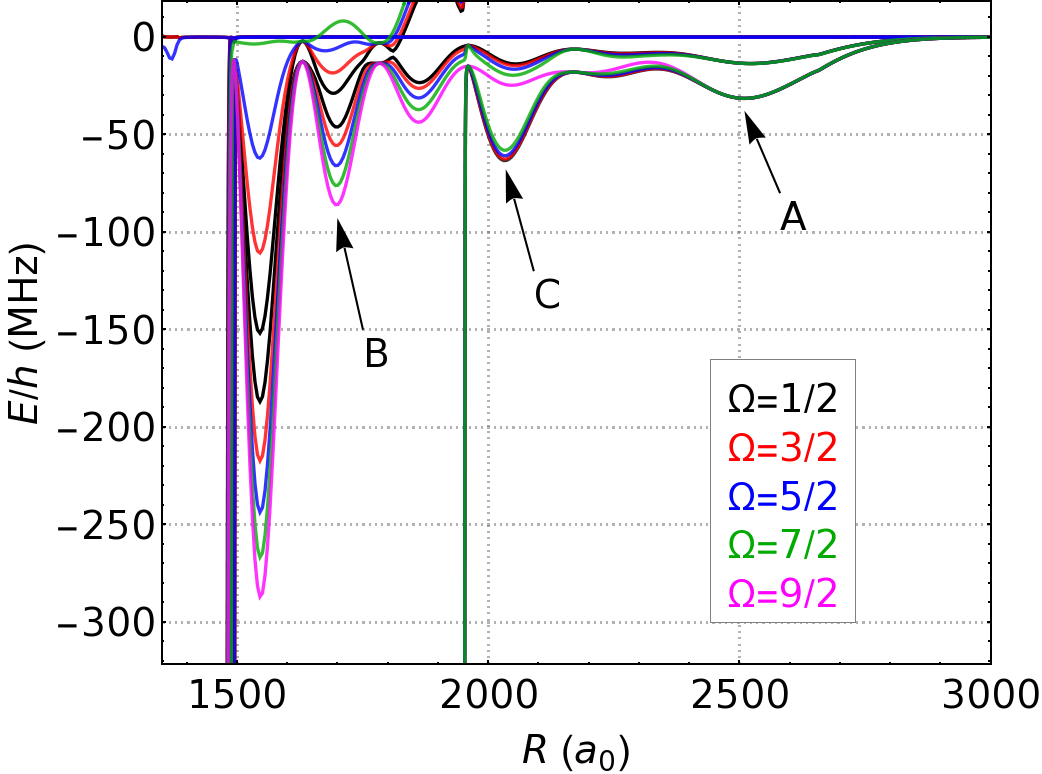}}
\caption{Calculated potential-energy curves of LRMs correlated to the atomic asymptote $40\,^2P_{3/2}-6\,^2S_{1/2}(F=4)$. The projection of the total angular momentum on the internuclear axis, $\Omega$, is indicated by the line color, see the legend. Features marked by letters are discussed in the text.} \label{fig:theo:pecs40P32F4overview}
\end{figure}

The resulting potential energy curves for LRMs correlated to the atomic asymptote $40\,^2P_{3/2}-6\,^2S_{1/2}(F=4)$ are depicted in Fig.~\ref{fig:theo:pecs40P32F4overview}. Caesium has a large nuclear spin of $i=7/2$ and, for this Rydberg state with $l=1,j=3/2$, $|\Omega|$ takes values in the range from $1/2$ to $11/2$. Note that $\Omega=11/2$ corresponds to the fully stretched configuration $m_i=7/2$, $m_j=3/2$ and $m_l=1$. Because the spatial wavefunction for a Rydberg state with $m_l=1$ has a node on the $z$-axis where the perturber is located, electronic states with $\Omega=11/2$ show only very weak Fermi-contact interactions~\cite{peperPhotodissociationLongrangeRydberg2020}. For all other values of $\Omega$, oscillatory potential-energy curves are observed. Close to the classical outer turning point of the Rydberg-electron orbit, around  $R=\num{2500}\,a_0$ and marked as region ``A'' in Fig.~\ref{fig:theo:pecs40P32F4overview}, the potential energy curves for different values of $\Omega$ are degenerate, splitting into two characteristic minima. The deeper one, with a dissociation energy of about $h\times\SI{30}{\mega\hertz}$, has been referred to as ``deep well''~\cite{andersonAngularmomentumCouplingsLongrange2014} or triplet-scattering-dominated well~\cite{sassmannshausenExperimentalCharacterizationSinglet2015}. The second, with a dissociation energy of about $h\times\SI{15}{\mega\hertz}$, has been referred to as ``shallow well''~\cite{andersonAngularmomentumCouplingsLongrange2014} or mixed-singlet-triplet-scattering well~\cite{sassmannshausenExperimentalCharacterizationSinglet2015}. The observed degeneracy results from the near-conservation of the angular momentum $\vec{N}=\vec{S}+\vec{i}$~\cite{deissObservationSpinorbitdependentElectron2020}, which in the case of the asymptote shown in Fig.~\ref{fig:theo:pecs40P32F4overview} yields the conserved quantum numbers $N=9/2$ (the deep well) and $N=7/2$ (the shallow well). 

This degeneracy is lifted by the spin-orbit interaction in the electron-caesium scattering, as explained in Ref.~\cite{deissObservationSpinorbitdependentElectron2020} and Appendix~\ref{sec:appendixSO}. The $P$-wave contribution to the Fermi-contact interaction~\eqref{eq:fermicontact} becomes significant around $R=\num{2000}\,a_0$, marked as region ``C'' in Fig.~\ref{fig:theo:pecs40P32F4overview}. Just below $R=\num{2000}\,a_0$, the semi-classical kinetic energy of the electron, see Eq.~\eqref{eq:semiclassk}, reaches the value of the $^3P_0$ shape resonance in the elastic electron-caesium scattering complex \cite{scheerExperimentalEvidenceThat1998,bahrimLowlying3P3S2000}. $P$-wave shape resonances cause a rapid change in the potential-energy curves of LRMs, visible as an abrupt drop of the potential-energy at internuclear separations just below the region ``C'' in Fig.~\ref{fig:theo:pecs40P32F4overview}, which result in very deep potential wells of so-called ``butterfly character''~\cite{hamiltonShaperesonanceinducedLongrangeMolecular2002}. In the region labeled ``B'' in Fig.~\ref{fig:theo:pecs40P32F4overview}, the degeneracy of states with different values of $\Omega$ is completely lifted and $N$ is no longer a good quantum number. 

\paragraph*{Vibrational levels}

The vibrational levels of the electronic states are obtained by the modified Milne phase-amplitude method on a grid spanning from $R_\mathrm{min}\sim\num{800}\,a_0$ to $R_\mathrm{max}\sim\num{3000}\,a_0$. WKB solutions for a plane wave are chosen as inner boundary conditions at $R_\mathrm{min} $~\cite{sidkyPhaseamplitudeMethodCalculating1999,peperHeteronuclearLongRangeRydberg2021}. Due to the open boundary conditions, bound-state solutions acquire a finite width $\Gamma$, given by the energy derivative of the scattering phase $\phi$ at the resonance energy $E_\mathrm{res}$~\cite{sidkyPhaseamplitudeMethodCalculating1999}:
\begin{equation}
    \left. \frac{\partial \phi}{\partial E}\right|_{E=E_{\mathrm{res}}}=\frac{2}{\Gamma}.
\end{equation}
For bound states (\textit{i.e.}, states mostly localised in between classically-forbidden regions), we interpret this width as resulting from tunneling through the barrier towards shorter internuclear distances with a time constant $\tau=1/(2 \pi\, \Gamma)$. Once an LRM has tunneled through this potential-energy barrier, it can decay very rapidly through vibronic couplings, either via a state-changing collision or autoionisation (see \textit{iv)} and \textit{v)} in Sec.~\ref{sec:introduction}, respectively)~\cite{schlagmullerUltracoldChemicalReactions2016}. We thus take $\tau$ as the lifetime of molecules in the vibrational level. Numerically,  the differential equations for phase and amplitude of the wavefunction are solved on a grid of binding energies, which is iteratively refined until the steps in the phase $\phi$ at $R_\mathrm{min}$ are less than \SI{0.2}{rad}.  

The calculated widths of the bound state resonances are strongly influenced by the width and height of the neigbouring potential barriers. However, we also find that the calculated width of a level depends to some extent on the shape of the potential-energy curve at shorter internuclear separations, where the electronic-structure calculations based on the Hamiltonian~\eqref{eq:isoelectronic} are less reliable and non-adiabatic contributions can become dominant~\cite{schlagmullerUltracoldChemicalReactions2016,durstNonadiabaticCouplingsStabilization2025}. To assess the variability of the calculated lifetimes, we replace the potential-energy curves at short internuclear separations with a smoothly decreasing function and systematically vary $R_\mathrm{min}$, which causes a periodic modulation of the calculated lifetimes. In the following, we report the center and width of this modulation as the value of the lifetime and its uncertainty, respectively. We note that the binding energies do not change significantly. 

\begin{figure}
\centering
\resizebox*{10cm}{!}{\includegraphics{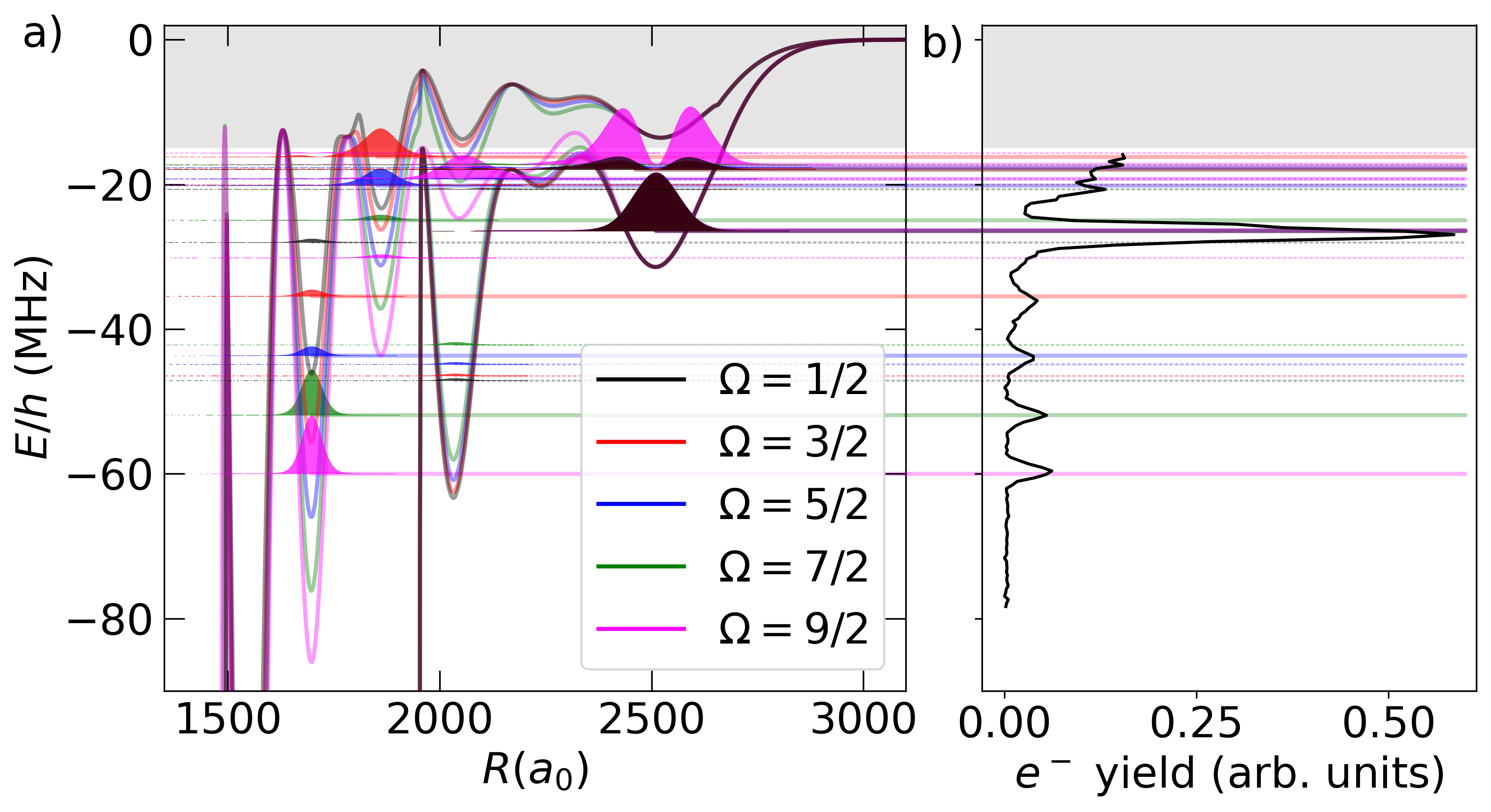}}
\caption{a) Potential energy curves of selected states from Fig.~\ref{fig:theo:pecs40P32F4overview} are shown on an enlarged scale. The vibrational levels of these states are illustrated by their probability densities, which overlay the curves. The baseline of each overlay indicates the binding energy of the vibrational level, and its area is proportional to the level’s lifetime. Levels with binding energies within the gray-shaded area are omitted from the figure for clarity. b) Late-PFI signal (see Sect.~\ref{sec:experiment} for details) for different detunings of the photoassociating UV laser from the transition $40\,^2P_{3/2}\leftarrow6\,^2S_{1/2}(F=4)$ and a photoassociation-pulse length of \SI{10}{\micro\second}. Before excitation, all atoms are optically pumped into the $6~^2S_{1/2}(F=4)$ state.
Calculated binding energies are shown as solid or dotted lines depending on whether their lifetimes are longer or shorter than \SI{1.2}{\micro\second}. Because the lines are drawn semi-transparently, overlapping resonances accumulate opacity and appear darker, as apparent for the $\Omega$-degenerate resonances at \SI{-26.4}{\mega\hertz}.
} \label{fig:theo:pecs40P32F4vibs}
\end{figure}

In Fig.~\ref{fig:theo:pecs40P32F4vibs}\,a), the vibrational levels of the states displayed in Fig.~\ref{fig:theo:pecs40P32F4overview} are illustrated by their vibrational wavefunctions. In the deep outermost potential well, the $v=0$ levels  have very narrow widths on the order of \SI{1}{\hertz}, hence their lifetimes are not limited by tunnelling and vibronic couplings.
The first excited vibrational state $v=1$ is long-lived only for $\Omega=9/2$, because the $^3P_0$ scattering channel does not contribute to the binding in this symmetry and thus the rapid drop of the potential-energy curve below $R=\num{2000}\,a_0$ is absent for this state. 

Vibrational levels bound in the inner wells around $R=\num{1700}\,a_0$, marked as region ``B'' in Fig.~\ref{fig:theo:pecs40P32F4overview}, exhibit a complete lifting of the $\Omega$-degeneracy. Because these nearly harmonic wells are very narrow, the vibrational frequencies are very large (comparable to the dissociation energies of the wells), and each well supports only a single long-lived vibrational level. The $v=0$ levels of the states with $\Omega=1/2, 3/2, \dots, 9/2$ form a nearly-even-spaced ladder with binding energies ranging from \SI{-28}{\mega\hertz} to \SI{-60}{\mega\hertz}, respectively. The calculated lifetimes increase with increasing binding energy, from \SI{1.0}{\micro\second} for the $v=0$ level of the $\Omega=1/2$ state to \SI{17}{\micro\second} for the $v=0$ level of the $\Omega=9/2$ state. This behaviour is governed by tunnelling through the surrounding potential barriers, with lifetimes set by the exponential dependence of the tunnelling rate on the barrier width and height.
In particular, the lifetime increase in region ``B’’  is explained primarily by the greater barrier width at $R=\num{1639}\,a_0$, and to a lesser extent by differences in the shape of the barrier at $R=\num{1500}\,a_0$, through both of which the particles have to tunnel. 

Vibrational levels bound in the intermediate wells around $R=\num{2050}\,a_0$, marked as region ``C'' in Fig.~\ref{fig:theo:pecs40P32F4overview}, are much shorter-lived than the levels bound at similar energies in region ``B''. This arises from the shape resonance in the $^3P_0$ scattering channel, which narrows the potential-energy barrier at $R=\num{2000}\,a_0$ and strongly enhances tunneling relative to the barrier at $R=\num{1639}\,a_0$.

Notably, the long-lived levels of the states with $\Omega<9/2$ have contributions from the $^1P_1$ scattering channel, which makes them promising candidates for creating ultracold Cs$_2$ molecules in highly-excited ion-pair states (also called ``heavy-Rydberg states''~\cite{reinholdHeavyRydbergStates2005}) through stimulated de-excitation of LRMs~\cite{peperFormationUltracoldIon2020,hummelUltracoldHeavyRydberg2020}.

\section{Experimental methods}\label{sec:experiment}

The experimental setup and procedures have been described previously~\cite{peperHeteronuclearLongRangeRydberg2021,peperRoleCoulombAntiblockade2023}, and here we revisit only the main features. A cloud of \SI{2e7} caesium atoms at a density of \SI{7e10}{\per\cubic\cm} and typical temperatures of \SI{40}{\micro \kelvin} is prepared in the $6\,^2S_{1/2} (F=4)$ ground state by laser cooling in a magneto-optical trap, followed by magnetic compression and optical molasses. At the end of the preparation stage, the atoms are optically pumped into the $F=4$ ground state with uncontrolled population of the near-degenerate Zeeman sublevels in a residual field of about \SI{10}{\milli G}. A modified ring dye laser system (Coherent 899-21), pumped by a frequency-doubled continuous-wave Nd:YVO4 laser, and a frequency-doubling unit (Coherent MBD 200) are used to produce frequency-tunable light for the excitation into Rydberg states at wavelengths around \SI{319}{\nano\meter}. A variable-offset-electronic-sideband locking scheme transfers the stability of a reference laser, stabilised via modulation-free saturated absorption spectroscopy in potassium vapor, to the dye-laser system. Residual frequency drifts are on the order of \SI{100}{\kHz\per\day}. An acousto-optic modulator is used to create short pulses of variable intensity and length which are applied to the atomic sample. Typical line widths for the excitation of ground-state atoms to Rydberg states with $n\sim40$ are on the order of \SI{1.7}{MHz}. During laser excitation, electric fields are actively compensated to better than \SI{20}{\milli\volt\per\centi\meter}~\cite{sassmannshausenHighresolutionSpectroscopyRydberg2013}. 

\begin{figure}
\centering
\resizebox*{7cm}{!}{\includegraphics{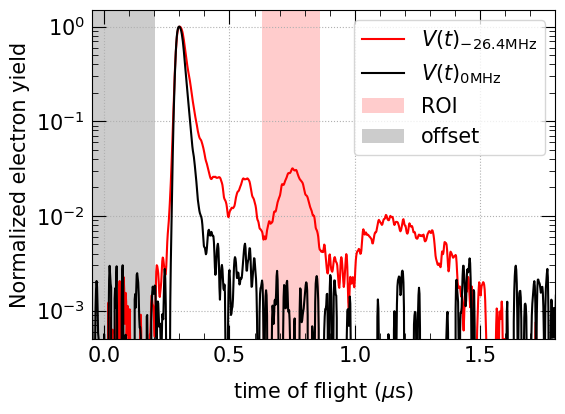}}
\caption{Comparison of electron time-of-flight traces recorded after exciting the ensemble of ground-state atoms at the atomic resonance $40\,^2P_{3/2}\leftarrow6\,^2S_{1/2}(F=4)$ ($V(t)_{\SI{0}{\mega\hertz}}$) and at a photoassociation resonance with a detuning of \SI{-26.4}{\mega\hertz} ($V(t)_{\SI{-26.4}{\mega\hertz}}$). The traces have been normalised to the same maximal amplitude and are drawn on a logarithmic scale. Shaded regions mark the time intervals in which the time-of-flight traces were integrated to obtain the late-PFI signal (red-shaded region) and the background-correction signal (gray-shaded region).} \label{fig:exp:toftraces_lrm_at}
\end{figure}

Excitation of Rydberg states is detected by pulsed-field ionisation (PFI) using a ramped electric field and subsequent detection of the resulting ions or electrons using a micro-channel-plate (MCP) detector~\cite{sassmannshausenHighresolutionSpectroscopyRydberg2013}. The current created at the anode of the MCP stack is amplified using a trans-impedance amplifier and recorded with a fast oscilloscope. We refer to the traces recorded by, and read out from, the oscilloscope as ``time-of-flight traces''. Exemplary time-of-flight traces of electrons detected after PFI are shown in Fig.~\ref{fig:exp:toftraces_lrm_at}. The black trace depicts the signal recorded after exciting isolated Rydberg atoms in the $40 {\hphantom{\,}^2P}_{3/2}$ state using UV pulses with very low intensity. A low-pass filter with a time constant of about \SI{1.2}{\micro\second}, inserted between the high-voltage-pulse switch and the electrodes, causes an exponentially rising electric field at the position of the Rydberg atoms. At about \SI{0.3}{\micro\second}, the strength of the electric field is sufficient to ionise the Rydberg atoms. The resulting electrons reach the detector almost instantaneously. Within \SI{0.2}{\micro\second}, all Rydberg atoms are ionised and no further electrons are detected. 

In a subsequent measurement, the frequency of the UV light is set to a detuning of \SI{-26.4}{\mega\hertz} below the transition frequency to the isolated Rydberg state, and the UV laser power is increased to \SI{100}{\milli\watt}. Under these conditions, long-lived LRMs are formed in the $v=0$ level of the outermost well of the $N=9/2$ states. The time-of-flight trace of the detected electron signal is depicted as red line in Fig.~\ref{fig:exp:toftraces_lrm_at}: after the first peak, identical to the one observed at the atomic resonance, the signal does not return to zero but decays much more slowly, exhibiting an oscillatory behavior with electrons arriving as late as \SI{1.5}{\micro\second} after the start of the ionisation-field ramp. In the following we call the fraction of electrons arriving after the first peak the ``late-PFI signal''. A related broadening of time-of-flight traces was reported for LRMs of strontium \cite{camargoLifetimesUltralongrangeStrontium2016,whalenLifetimesUltralongrangeStrontium2017}, where both state-changing collisions and LRM-specific field-ionisation dynamics were discussed as possible explanation. 

We perform rf-induced photodissociation of LRMs to probe the character of the Rydberg sample immediately before field ionisation~\cite{peperPhotodissociationLongrangeRydberg2020}. Specifically, we apply a \SI{10}{\micro\second} UV light pulse, followed by a \SI{10}{\micro\second} RF pulse. When the RF frequency is resonant with the transition from the LRM with a binding energy of \SI{-26.4}{\mega\hertz} to the $39^2D_{5/2} -  6^2S_{1/2}(F=4)$ continuum, the late-PFI signal returns to background, confirming that the late-PFI signal originates from LRMs present immediately before the PFI ramp. This excludes black-body induced transitions or state-changing collisions before application of the ionisation field as explanation for the late-PFI signal. The exact mechanism causing the late-PFI signal is still under investigation, however our observations are consistent with up to \SI{20}{\percent} of LRMs field-ionising along a more diabatic pathway~\cite{gallagherRydbergAtoms1994} than the isolated atom. 

As discussed in previous work~\cite{peperRoleCoulombAntiblockade2023}, ions present in the sample during photoassociation facilitate the excitation of isolated Rydberg atoms at large detunings from the atomic resonance, a mechanism called ``Coulomb anti-blockade''~\cite{boundsCoulombAntiblockadeRydberg2019}. Decay of an LRM through associative ionisation creates such ions in the sample, causing the facilitated excitation of additional Rydberg atoms which complicate the analysis of the ionisation signals~\cite{peperRoleCoulombAntiblockade2023}. Fortunately, these additionally excited Rydberg atoms do not exhibit the late-PFI ionisation behavior discussed above. The late-PFI signal is therefore an unambiguous measure of the number of LRMs in the excitation region immediately before the PFI field is applied and is ideally suited to study the lifetime of LRMs.

We extract the late-PFI signal from the time-of-flight traces by integrating the electron yield in the regions of interests (ROIs) shaded in red and gray in Fig.~\ref{fig:exp:toftraces_lrm_at}. The latter integral is used to correct for variations of the signal's offset due to electrical noise. The resulting late-PFI signal is shown in Fig.~\ref{fig:theo:pecs40P32F4vibs}\,b) when the detuning of the UV laser pulse is scanned around the transition $40\,^2P_{3/2}\leftarrow6\,^2S_{1/2}(F=4)$. The largest peak at a detuning around \SI{-26.4}{\mega\hertz} corresponds to the photoassociation of LRMs in the $v=0$ level of the outermost well of the $N=9/2$ states. We observe four additional peaks for larger detunings between \SI{-60}{\mega\hertz} and \SI{-36}{\mega\hertz}. Overlaid onto the experimental spectrum are the binding energies resulting from the calculations explained in Sec.~\ref{sec:theory}, grouped into levels with calculated decay times larger and smaller than a chosen value of \SI{1.2}{\micro\second}. The observed resonances in the late-PFI spectrum agree very well with the predicted binding energies of long-lived LRMs, and we confidently assign the resonances at \SI{-60}{\mega\hertz}, \SI{-52}{\mega\hertz}, \SI{-44}{\mega\hertz}, and \SI{-36}{\mega\hertz} to the photoassociation of LRMs in the deep wells at $R=\num{1700}\,a_0$ (see Fig.~\ref{fig:theo:pecs40P32F4vibs}\,a)) with $\Omega=9/2$, $\Omega=7/2$, $\Omega=5/2$, and $\Omega=3/2$, respectively. We refer to these LRMs as the ones bound in the ``inner well''.  

\section{Lifetime measurements}\label{sec:lifetimes}

We measure the lifetimes of LRMs formed in different states by recording the late-PFI signal for typically 20 different delays between the end of the UV pulse and the rise of the PFI field, integrating in the oscilloscope over 100 experimental repetitions and repeating the measurement at each delay between 8 and 20 times. We determine the uncertainty of the integral for each delay by bootstrapping the sample typically hundred times~\cite{efronBootstrapMethodsAnother1992}. We eventually extract the decay time constant by fitting the integrals for different delays with an exponentially decaying curve in a nonlinear regression analysis, weighted by the uncertainty of each integral. We extract the uncertainty of the decay constant from the covariance matrix of the regression analysis. We have explored different approaches to correct for electronic offsets of the time-of-flight traces, such as fitting the recorded time-of-flight traces with different empirical functions, and obtained decay time constants which agree within their extracted uncertainties. We thus consider our analysis to be robust against details of the signal extraction. 

The lifetime of LRMs formed in the $v=0$ level of the outermost well at a detuning of \SI{-26.4}{\mega\hertz} is found to be \SI{58(3)} {\micro\second}, in good agreement with the calculated lifetime of the parent Rydberg state $40^2P_{3/2}$ in the presence of black-body radiation (BBR) at $T=\SI{300}{\kelvin}$, $\tau_\mathrm{calc}=\SI{57.3}{\micro\second}$~\cite{sibalicARCOpensourceLibrary2017,robertsonARC30Expanded2021a}. We note that the lifetime of the parent atom is limited by BBR-induced transitions, because spontaneous decay to the ground state is strongly suppressed for caesium $n^2P_J$ states due to a Cooper minimum in the photoionisation cross section of caesium~\cite{raimondLaserMeasurementIntensity1978}. 
We find a slightly longer decay time of \SI{65(2)}{\micro\second} if we analyse the decay of the full integral of the time-of-flight traces (see Fig.~\ref{fig:exp:toftraces_lrm_at}), which is very close to the decay time we extract for the decay of the integrals recorded after the excitation of isolated $40^2P_{3/2}$ Rydberg atoms at the transition $40\,^2P_{3/2}\leftarrow6\,^2S_{1/2}(F=4)$, \SI{67(2)}{\micro\second}. The observation that the latter significantly exceeds $\tau_\mathrm{calc}$ can be attributed to the fact that BBR-induced transitions mostly populate neighboring Rydberg levels which have similar field-ionisation thresholds and which thus do not lead to a reduction of the integrated electron yield.

\begin{figure}
\centering
\resizebox*{7cm}{!}{\includegraphics{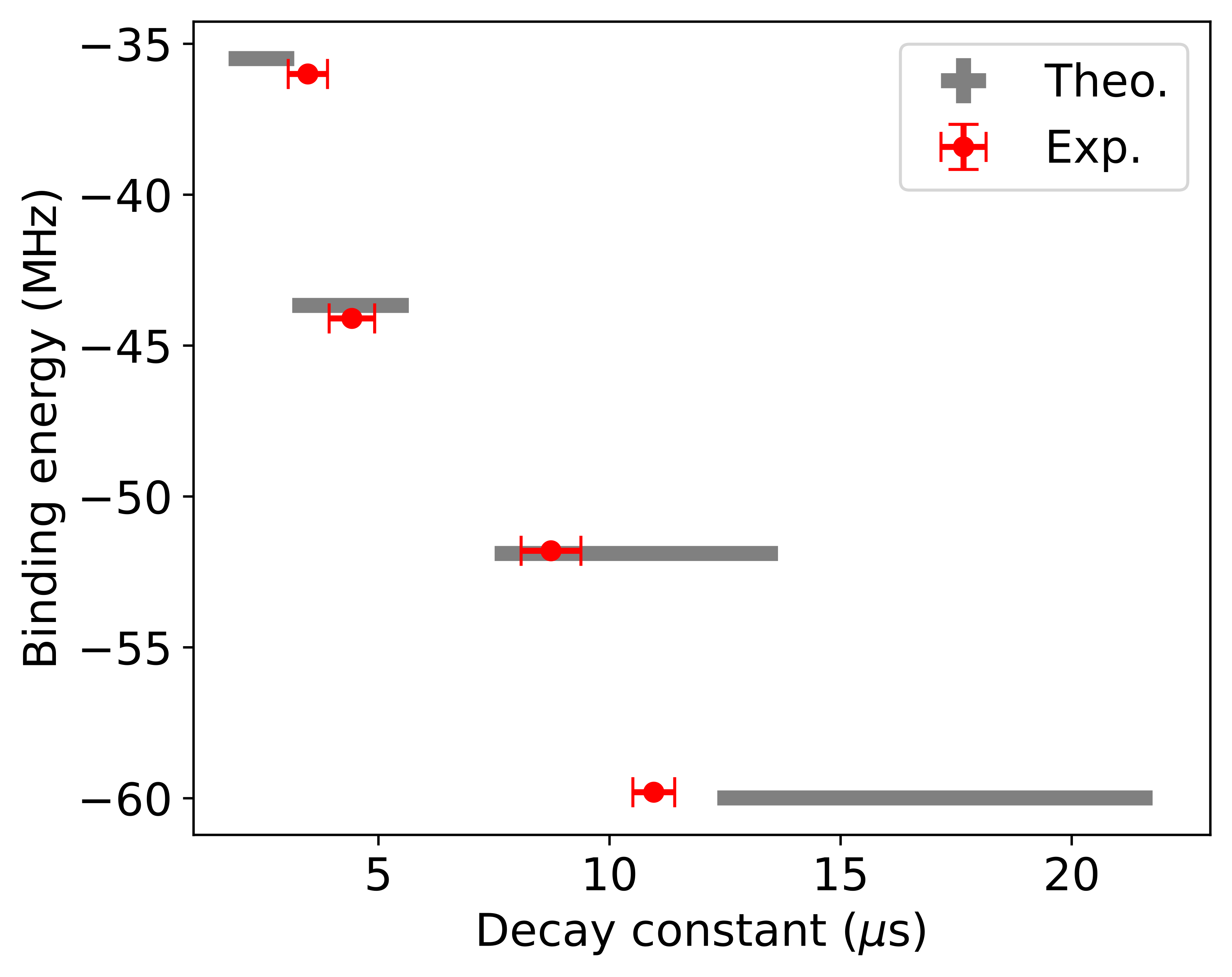}}
\caption{Comparison of decay constants and binding energies determined experimentally (red dots with error bars) and theoretically (gray bands) for the four photoassociation resonances with binding energies larger than \SI{30}{\mega\hertz} (compare Fig.~\ref{fig:theo:pecs40P32F4vibs}\,b)). The gray bands indicate the range of decay constants compatible with our calculations, see Sec.~\ref{sec:theory} for details.} \label{fig:disc:comparison_lifetimes}
\end{figure}

In comparison to these lifetimes, the lifetimes of LRMs formed at larger detunings are significantly reduced, as can be seen in Fig.~\ref{fig:disc:comparison_lifetimes}. The measured lifetimes are in good agreement with the calculated lifetimes of LRMs bound in the inner well and notably exhibit the same systematic dependence on the binding energy, lending further support to the assignment of the observed photoassociation resonances to the calculated vibronic states. In order to identify the dominant decay mechanism responsible for the reduced lifetime of the LRMs formed in the inner wells, we change the setup to detect the positively-charged ions arriving at our detector after photoassociation of LRMs. The decay product of associative ionisation is a stable Cs$_2^+$ ion. It can be easily discerned by its arrival time at the detector, which is delayed compared to the one of Cs$^+$ ions because of the larger mass. We observe Cs$_2^+$ ions for all LRMs bound in the inner well and find a systematic increase in the number of detected Cs$_2^+$ ions for LRMs with a shorter lifetime. We cannot make more quantitative statements because of the poorly specified mass- and velocity-dependent sensitivity of our MCP detector, but the observed strength of the Cs$_2^+$ signal is compatible with a large fraction of the LRMs studied here decaying through associative ionisation. 

Alternatively, LRMs can decay through state-changing collisions, which have been studied in detail in rubidium~\cite{schlagmullerUltracoldChemicalReactions2016,geppertDiffusivelikeRedistributionStatechanging2021}. The products of state-changing collisions are Rydberg atoms in neighboring, lower-lying Rydberg states which have a higher field-ionisation threshold. These decay products thus ionise later in the rising electric field ramp applied to the sample, and the produced electrons or ions arrive later at the detector. We note that the decay products are predominantly formed in Rydberg states with a change in effective quantum number $\Delta n^*<1$~\cite{geppertDiffusivelikeRedistributionStatechanging2021}, and that the resulting charges would arrive still much earlier than the late-PFI signals discussed above (see Fig.~\ref{fig:exp:toftraces_lrm_at}). Indeed, we observe a slight broadening of the main field-ionisation peak (see Fig.~\ref{fig:exp:toftraces_lrm_at}), but further investigations are necessary to quantitatively determine the branching ratio between the two decay processes, associative ionisation and state-changing collisions.

\section{Discussion}

Previous studies have reported a strong dependence of the lifetime of LRMs on the density of ground-state atoms~\cite{butscherLifetimesUltralongrangeRydberg2011,niederprumGiantCrossSection2015,baiDissociationUltracoldCesium2023} where mostly LRMs formed in the outermost wells were studied. In our experiments, the lifetime of LRMs in the $v=0$ level of the outermost well, with a binding energy of about \SI{-26.4}{\mega\hertz}, was found to be in agreement with a lifetime limited by BBR-induced transitions. The much shorter lifetimes found here for LRMs formed in the inner well, which have a smaller geometric cross section (see Fig.~\ref{fig:theo:pecs40P32F4vibs}\,a)), can thus not be significantly influenced by collisions with ground-state atoms. The observed lifetimes are in good agreement with our theoretical model which attributes the reduction in lifetime to tunneling through a potential-energy barrier and subsequent coupling to short-range intramolecular decay channels.

The binding energy of the LRMs formed in the inner well, marked as region B in Fig.~\ref{fig:theo:pecs40P32F4overview}, is dominated by the spin-orbit interaction of the Cs$^-$ collision complex, which lifts the degeneracy of the states with different values of $\Omega$ (see Appendix~\ref{sec:appendixSO}). Both the observed binding energy and the lifetime are greatly reduced as $\Omega$ goes from $9/2$ to $3/2$.  We thus argue that it is the spin-orbit interaction of the electron-caesium scattering complex which is responsible for the reduced lifetimes of the observed LRMs. This builds an interesting bridge to the spin-dependent autoionisation dynamics of molecules in Rydberg states where the ion core is a molecular ion with spin-orbit and hyperfine interactions~\cite{lefebvre-brionTheoreticalStudySpinorbit1985,wornerRoleSpinsMolecular2007}. LRMs bound in wells at even shorter internuclear distances would provide an interesting intermediate regime, describable either as a molecular Rydberg state with a vibrational highly excited ionic core or as an LRM.

The vibronic states observed and assigned in this work are promising candidates for creating ultracold molecules in highly-excited ion-pair states, which offer a route towards the preparation of ultracold anions and strongly correlated pair plasmas~\cite{peperFormationUltracoldIon2020,hummelUltracoldHeavyRydberg2020}.

\section*{Acknowledgement(s)}

We thank Ali-Dzhan Ali and Harsh Mishra for their contributions to the theoretical models employed in this work. We thank Frédéric Merkt, Matt Eiles, Francis Robicheaux, Chris Greene, Hossein Sadeghpour, Seth Rittenhouse, Daniel Vrinceanu, Thomas Killian, and Shuhei Yoshida for fruitful discussions on the field-ionisation dynamics of LRMs.

\section*{Disclosure statement}

The authors declare no potential conflicts of interest with respect to the research, authorship, or publication of this article.

\section*{Data availability statement}

The data, from which the figures and findings of this study were drawn, are available on request from the corresponding author, JD. 

\section*{Funding}

This work was funded by the Deutsche Forschungsgemeinschaft (DFG, German
Research Foundation) - 428456632 under SPP 1929 (GiRyd). 

\bibliographystyle{tfo}

\appendix

\section{Coupling of angular momenta}
\label{sec:couplingScheme}

\begin{figure}[bh]
\centering
\resizebox*{7cm}{!}{\includegraphics{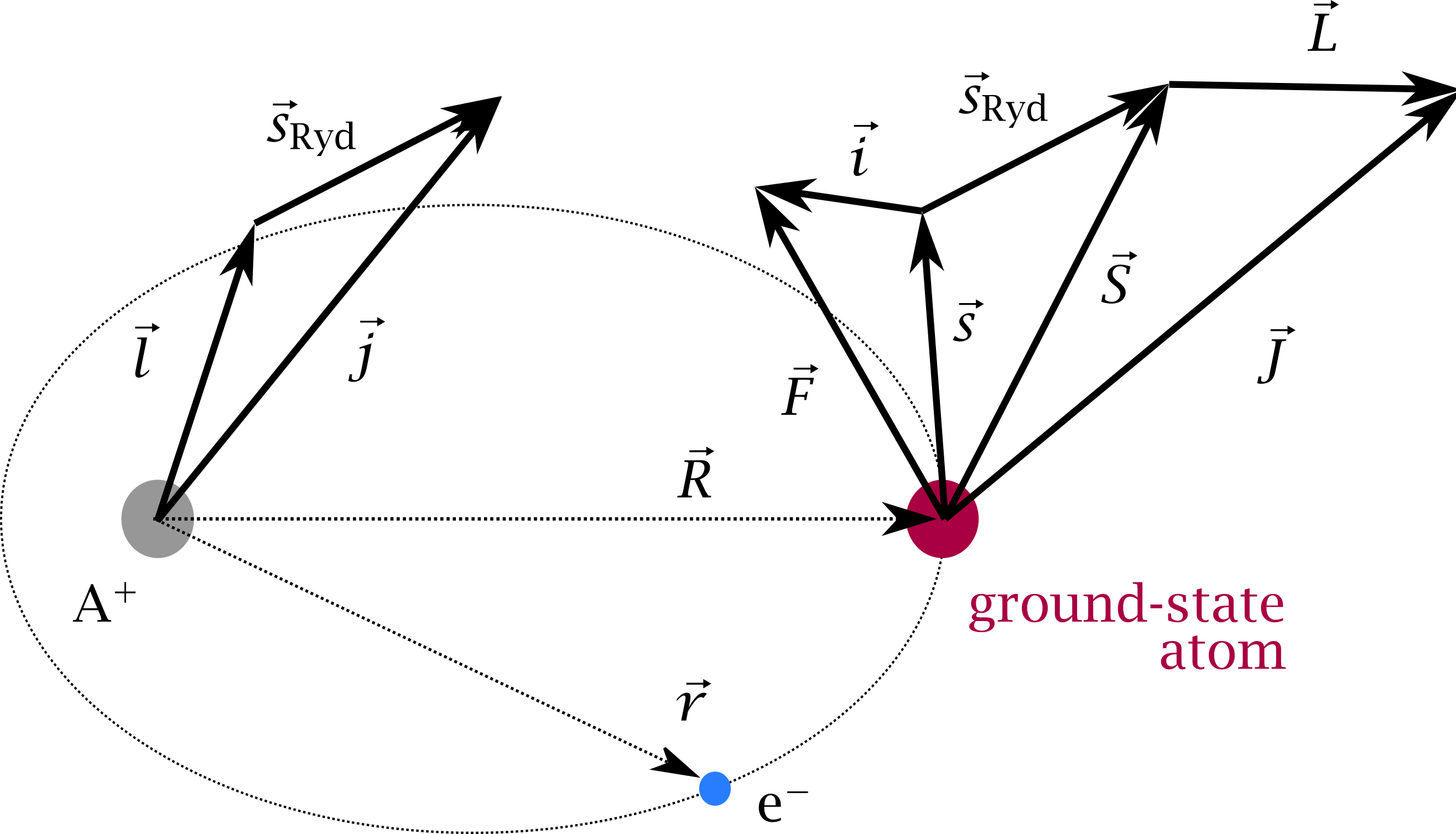}}
\caption{Schematic illustration of the couplings of angular momenta in a long-range Rydberg molecule composed of a ground-state atom GS (red circle) and a Rydberg atom with Rydberg ion core A$^+$ (gray circle) and electron $e^-$ (blue circle). The molecular axis is along the vector $\vec{R}$ from $A^+$ to GS, the electron's coordinate $\vec{r}$ is defined relative to $A^+$. The electron's orbital angular momentum $\vec{l}$ and its spin $\vec{s}_\mathrm{Ryd}$ couple to the Rydberg electron's total angular momentum $\vec{j}$. The spin $\vec{s}_\mathrm{Ryd}$ couples to the electronic spin of GS, $\vec{s}$, to form the total electronic spin $\vec{S}$. The orbital angular momentum of $e^-$ with respect to GS, $\vec{L}$, and $\vec{S}$ couple to form the total angular momentum $\vec{J}$ of the $e^-$-GS scattering state. The nuclear spin of GS, $\vec{i}$, couples with $\vec{s}$ to form $\vec{F}$, the total angular momentum including nuclear spin of GS. Figure and caption adapted from Figure 5.1 of Ref.~\cite{peperPrecisionSpectroscopyUltracold2020}.} \label{fig:appendix:couplingAngularMomenta}
\end{figure}

The coupling of angular momenta employed in the calculation of potential-energy curves (see section~\ref{sec:theory}) follows the coupling scheme presented by Eiles and Greene~\cite{eilesHamiltonianInclusionSpin2017}. It is illustrated schematically in Fig.~\ref{fig:appendix:couplingAngularMomenta}: the state of the Rydberg electron is described in the coordinate system centred around the Rydberg atom, with quantum numbers $l$, $s_\mathrm{ryd}$, $j$, and $m_j$. The elastic-scattering interaction between the Rydberg electron and the ground-state atom is described in the coordinate system centred around the ground-state atom with quantum numbers $L$, $S$, $J$, and $M_J$. A frame-transformation matrix connects the two coordinate systems for numerical computations. The hyperfine interaction of the ground-state atom is added separately, yielding quantum numbers $i$, $s$, $F$, and $m_F$ for the nuclear spin, the electronic spin, the total angular momentum of the ground-state atom, and its projection on the internuclear axis, respectively.

\section{Origin of $\Omega$-splitting}
\label{sec:appendixSO}
We follow the work of Dei{\ss} \textit{et al.}~\cite{deissObservationSpinorbitdependentElectron2020} to investigate the contributions of spin-orbit coupling in the electron-caesium scattering to the binding energy of the LRMs studied in this work. To this end, two parameters are introduced, $\lambda_1$ and $\lambda_2$, which parameterise the scattering phase shifts in the following way: for $\lambda_1=0$, singlet and triplet scattering interactions are identical so that there is no exchange interaction in the Cs$^-$ collision complex. For $\lambda_1=1$, the singlet and triplet scattering phase shifts are set to the values obtained as described in Sec.~\ref{sec:theory}. The parameter $\lambda_2$ influences the strength of the spin-orbit coupling, which requires $L=S=1$: for $\lambda_2=0$, the $^3P_J$ phase shifts are identical, $^3P_J={^3P}_1$ for all $J$. For $\lambda_2=1$, the phase shifts of the three spin-orbit components are the ones obtained through the theoretical model (see  Sec.~\ref{sec:theory}). By calculating the binding energies with the help of Hamiltonian~\eqref{eq:isoelectronic} while varying the parameters from $\lambda_1=0,\lambda_2=0$ to $\lambda_1=1,\lambda_2=0$, and from $\lambda_1=1,\lambda_2=0$ to $\lambda_1=1,\lambda_2=1$, the contributions from the exchange interaction and the spin-orbit interaction, respectively, of the Cs$^-$ collision complex can be assessed. 

\begin{figure}
\centering
\resizebox*{9cm}{!}{\includegraphics{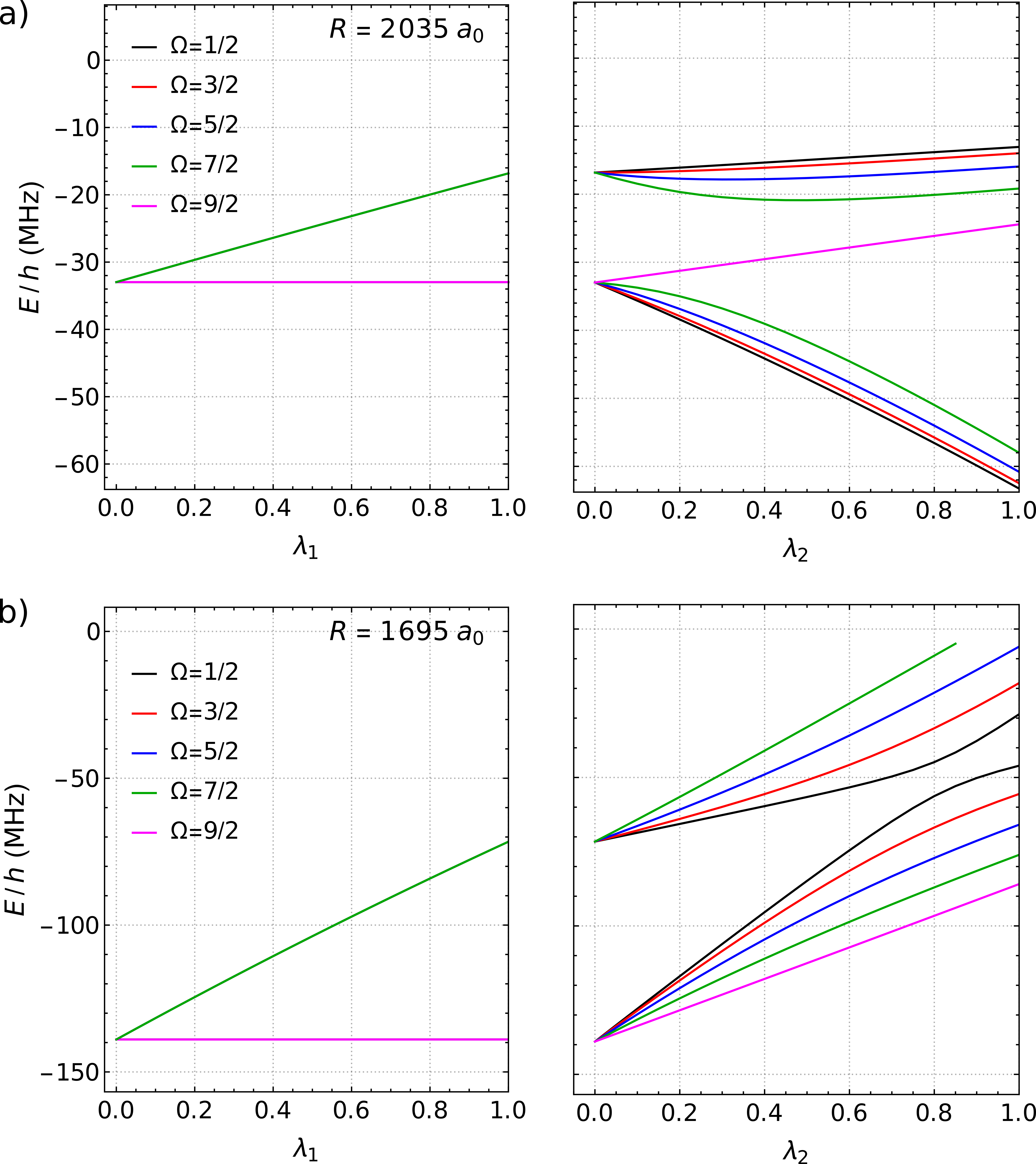}}
\caption{Analysis of the contributions from the exchange interaction (controlled through $\lambda_1$) and the spin-orbit coupling (controlled through $\lambda_2$) in the Cs$^-$ scattering complex to the electronic energies $E$ of LRMs at a) $R=\num{2035}\,a_0$, marked as region C in Fig.~\ref{fig:theo:pecs40P32F4overview}, and b) the internuclear separation of the inner well ($R=\num{1695}\,a_0$), marked as region B in Fig.~\ref{fig:theo:pecs40P32F4overview}. Energies are given with respect to the atomic asymptote $40^2P_{3/2}-6^2S_{1/2}(F=4)$. See text and Ref.~ \cite{deissObservationSpinorbitdependentElectron2020} for details.} \label{fig:appendix:roleofSO}
\end{figure}

For $\lambda_1=0,\lambda_2=0$, the states with the approximate quantum numbers $N=7/2$ and $N=9/2$ are degenerate, as can be seen in Fig.~\ref{fig:appendix:roleofSO} where the potential energies at two specific internuclear separations in the regions labeled B and C in Fig.~\ref{fig:theo:pecs40P32F4overview} are plotted as function of the parameters $\lambda_1$ and $\lambda_2$. When increasing  $\lambda_1$ to 1, the degeneracy in $N$ is lifted, resulting in two groups of states which correspond to the triplet- and mixed-singlet-triplet-scattering states discussed in earlier work~\cite{sassmannshausenExperimentalCharacterizationSinglet2015}. The states remain degenerate in $\Omega$. When increasing $\lambda_2$ from 0 to 1, this degeneracy is also lifted. While $N$ remains an approximate quantum number at the internuclear separation $R=\num{2035}\,a_0$, the spin-orbit induced splitting of the potential-energy curves exceeds the contribution from the exchange-interaction and $\Omega$ is the only remaining good quantum number describing the LRMs bound in the inner well at $R=\num{1695}\,a_0$.

\end{document}